\documentclass[prd,amssymb,twocolumn,10pt]{revtex4}
\usepackage{amsmath,amsfonts,amssymb}
\usepackage{verbatim,graphicx,float}

\begin{document}

\title{Exact solutions for Big Bounce in loop quantum cosmology}

\author{Jakub Mielczarek}
\email{jakubm@poczta.onet.pl}
\affiliation{\it Astronomical Observatory, Jagiellonian University, 30-244
Krak\'ow, Orla 171, Poland}
\affiliation{\it The Niels Bohr Institute, Copenhagen University, Blegdamsvej 17, 
DK-2100 Copenhagen, Denmark}

\author{Tomasz Stachowiak}
\email{toms@oa.uj.edu.pl}
\affiliation{\it Astronomical Observatory, Jagiellonian University, 30-244
Krak\'ow, Orla 171, Poland}

\author{Marek Szyd{\l}owski}
\email{uoszydlo@cyf-kr.edu.pl}
\affiliation{Department of Theoretical Physics, 
Catholic University of Lublin, Al. Rac{\l}awickie 14, 20-950 Lublin, Poland}
\affiliation{Marc Kac Complex Systems Research Centre, Jagiellonian University,
Reymonta 4, 30-059 Krak{\'o}w, Poland}

%\date{\today}

\begin{abstract}
In this paper we study the flat ($k=0$) cosmological FRW model with holonomy corrections
of Loop Quantum Gravity. The considered universe contains a
massless scalar field and the cosmological constant $\Lambda$. We find analytical 
solutions for this model in different configurations and investigate  
its dynamical behaviour in the whole phase space. We show the explicit 
influence of $\Lambda$ on the qualitative and quantitative character 
of solutions. Even in the case of positive $\Lambda$ the oscillating 
solutions without the initial and final singularity appear 
as a generic case for some quantisation schemes.  
\end{abstract}

\maketitle

\section{Introduction} \label{sec:intro}

In recent years Loop Quantum Cosmology (LQC) has inspired
realisation of the cosmological scenario in which the initial 
singularity is replaced by the bounce. In this picture, the
Universe is initially in the contracting phase, reaches the 
minimal, nonzero volume and, thanks to quantum repulsion,
evolves toward the expanding phase. Such a scenario has been 
extensively studied with use of the numerical methods 
\cite{Singh:2006im,Ashtekar:2006uz}.  
However, as it was shown for example in \cite{Stachowiak:2006uh}
exact solutions for bouncing universe with dust and cosmological 
constant can be found. The aim of the present paper is to show 
that analytical solutions can also be obtained for
the bouncing models arising from LQC. The main 
advantage of such exact solutions is that they allow for
investigations in whole ranges of the parameter domains. 

In this paper we consider the flat FRW model with a free
scalar field and with the cosmological constant. Quantum effects are
introduced in terms of correction to the classical theory. Generally one
considers
two types of of quantum correction: correction from inverse volume and 
holonomy corrections.
The leading effect of the volume corrections is the appearance of the super-inflationary phase.
The effect of holonomy corrections, on the other hand, is the appearance of a
bounce instead of singularity. The aim of this paper is to investigate analytically
these effects in a flat FRW model. That is to say, we neglect corrections
from inverse volume, these effects however, has been extensively studied
elsewhere. 
Moreover, these two types of corrections are not equally important 
in the same regimes. The inverse volume corrections are mainly important 
for small values of the scale factor, whereas holonomy corrections are 
mainly important for large values of the Hubble parameter. In other words,
when the minimal scale factor (during the bounce) is large enough, 
the effects of inverse volume corrections can be neglected.

The flat FRW model in the Loop Quantum Cosmology 
has been first investigated in the pioneer works of Bojowald 
\cite{Bojowald:2002gz,Bojowald:2003md}
and later improved in the works of Ashtekar, Paw{\l}owski and Singh 
\cite{Ashtekar:2006rx,Ashtekar:2006uz,Ashtekar:2006wn}.    
Bojowald's original description of the quantum universe in currently
explored in the number of works and regarded as a parallel line 
of research \cite{Bojowald:2006gr,Bojowald:2008pu}.
In the present paper, we restrict ourselves to the 
flat FRW models arising in the framework proposed by Ashtekar and co-workers.   
Beside the flat models this approach has also been applied to 
the FRW  $k= \pm 1$ models in
\cite{Ashtekar:2006es,Vandersloot:2006ws,Szulc:2006ep,Szulc:2007uk}  
and Bianchi I in \cite{Chiou:2007mg,Chiou:2007dn}.
In these models the unambiguity in the
choice of the elementary area for the holonomy corrections appear.
In the present paper we consider two kind of approaches to this problem: the so
called $\bar{\mu}-$scheme and $\mu_0-$scheme (for a more detailed description
see Appendix \ref{Appendix1}).
We find analytical solutions for the considered models in these two schemes.   

The Hamiltonian of the considered model is given by   
\begin{equation}
H_{\text{eff}} =  - \frac{3}{8 \pi G \gamma^2} \sqrt{|p|} \left[ \frac{ \sin \left( \bar{\mu} c
\right) }{\bar{\mu}}   \right]^2 
+ \frac{1}{2} \frac{ p_{\phi}^2}{  {|p|}^{3/2} } +{|p|}^{3/2}\frac{\Lambda}{8\pi G}.
\label{model}
\end{equation}
In Appendix \ref{Appendix1} we show the derivation of this Hamiltonian in the 
Loop Quantum Gravity setting. The canonical variables for the gravitational field 
are $(c,p)$  and for the scalar field $(\phi,p_{\phi})$. The canonical 
variables for the gravitational field can be expressed
in terms of the standard FRW variables  $(c,|p|)=(\gamma \dot{a} V^{1/3}_0,a^2V^{2/3}_0 )$.
Where the factor $\gamma$ is called Barbero-Immirzi parameter and is a constant
of the theory, and $ V_0$ is the
volume of the fiducial cell. The volume $ V_0$ is just a scaling factor and can be chosen
arbitrarily in the domain $V_0\in \mathbb{R}_+$. Since $p$ is the more natural
variable than $a$ here, we present mostly $p(t)$ in the figures. $a$ is always
the positive square root of $p$ so the shape of the graphs would be essentially
the same when drawn with $a$. The equations of
motions can now be derived with the use of the Hamilton equation
\begin{equation}
\dot{f} = \{f,H_{\text{eff}}  \}
\label{Hameq}
\end{equation}
where the Poisson bracket is defined as follows
\begin{eqnarray}
\{f,g\} &=& \frac{8 \pi G \gamma }{3}
\left[\frac{\partial f}{\partial c}\frac{\partial g}{\partial p}- 
\frac{\partial f}{\partial p}\frac{\partial g}{\partial c}  \right]   \nonumber \\
&+& \left[\frac{\partial f}{\partial \phi}\frac{\partial g}{\partial p_{\phi} }- 
\frac{\partial f}{\partial p_{\phi}}\frac{\partial g}{\partial \phi}  \right]. 
\end{eqnarray}
From this definition we can immediately retrieve the elementary brackets
\begin{equation}
\{c,p \} = \frac{8 \pi G \gamma }{3} \ \ \text{and} \ \  \{\phi,p_{\phi} \} = 1. 
\end{equation}
With use of the Hamiltonian (\ref{model}) and equation (\ref{Hameq}) we obtain
equations of motion for the canonical variables
\begin{eqnarray}
\dot{p} &=& \frac{2}{\gamma} \frac{ \sqrt{|p|} }{\bar{\mu} }  \sin \left( \bar{\mu} c 
\right) \cos \left( \bar{\mu} c  \right), 
\nonumber \\
\dot{c} &=&  - \frac{1}{\gamma} \frac{\partial}{\partial p} \left\{ \sqrt{|p|} 
\left[ \frac{ \sin \left( \bar{\mu} c \right)  }{\bar{\mu}}   \right]^2   \right\} \nonumber \\ 
  &-&\text{sgn}(p)\frac{\kappa \gamma}{4}  \frac{p_{\phi}^2}{{|p|}^{5/2}} + \text{sgn}(p)   \frac{\Lambda \gamma }{2}\sqrt{|p|}, 
\nonumber  \\
\dot{\phi} &=&      {|p|}^{-3/2} p_{\phi},  \nonumber  \\
\dot{p_{\phi}} &=&   0,
\label{equations}
\end{eqnarray}
where $\kappa=8\pi G$.
The Hamiltonian constraint $H_{\text{eff}} =0$ implies
\begin{equation}
\frac{1}{\gamma^2 |p|}  \left[ \frac{ \sin \left( \bar{\mu} c \right)  }{\bar{\mu}}   \right]^2 =
\frac{\kappa }{3} \frac{1}{2} \frac{p_{\phi}^2}{{|p|}^3} +\frac{\Lambda}{3}.
\label{constraint}
\end{equation}

The variable $\bar{\mu}$ corresponds to the dimensionless length of the edge of the
elementary loop and can be written in the general form
\begin{equation}
\bar{\mu}(p) = \xi {|p|}^n
\label{correction}
\end{equation}
where $-1/2 \leq n \leq 0$ and $\xi$ is a constant $\xi>0$ (this comes from the 
fact that $\mu$ is positively defined). The choice of $n$
and $\xi$ depends on the particular scheme in the holonomy corrections.
In particular, boundary values correspond to the cases when $\bar{\mu}$ is
the physical distance ($n=-1/2$, $\bar{\mu}-$scheme) and coordinate distance
($n=0$, $\mu_0-$scheme).
However, the $n=0$ case does not lead to the correct classical limit.
When $n>0$, the classical limit can not be recovered either. Only 
for negative values of $n$ is the classical limit $p\rightarrow \infty$ 
correctly recovered
\begin{equation}   
\lim_{p\rightarrow \infty} \frac{ \sin \left( \bar{\mu}(p) c \right)  }{\bar{\mu}(p)}  = c. 
\end{equation}

Strict motivation of the domain of the parameter $n$ comes form 
the investigation of the lattice states \cite{Bojowald:2006qu}. The number of the 
lattice blocks is expressed as $\mathcal{N} = V_0/l_0^3 $ where $l_0$ is the
average length of the lattice edge. This value is connected to 
the earlier introduced length $\bar{\mu}$, namely $\mathcal{N} = \bar{\mu}^{-3}(p)$.
During evolution the increase of the total volume is due to 
the increase of the spin labels on the graph edges or due to 
the increase of the number of vortices. In this 
former case the number of the lattice blocks is constant 
during evolution, $\mathcal{N} = \text{const}$.
Otherwise, when the spin labels do not change, the number 
of vortices scales with the volume, $\mathcal{N} \propto |p|^{3/2}$.
The physical evolution correspond to something 
in the middle, it means the power index lies in the range $[0,3/2]$. 
Applying the definition of $\mathcal{N}$ we
see that the considered boundary values translate to the domain of 
$n$ introduced earlier, $n\in [-1/2,0]$. 
More detailed investigation of the considered ambiguities 
can be found in the papers \cite{Bojowald:2008pu,Bojowald:2007yy}
 and in the Appendix \ref{Appendix1}.

Combining equations (\ref{constraint}), (\ref{correction}) and the first one from the set of equations (\ref{equations})
we obtain 
\begin{eqnarray}
\left(\frac{dp}{dt} \right)^2 &=& \Omega_{\text{I}} {|p|}^{-1}+   \Omega_{\text{II}}{|p|}^2 - \Omega_{\text{III}}{|p|}^{2n-3}  \nonumber \\
                              &-&  \Omega_{\text{IV}}{|p|}^{2n}
                              -\Omega_{\text{V}}{|p|}^{2n+3},
\label{mainequation}
\end{eqnarray}
where new parameters are defined as follow 
\begin{eqnarray}
\Omega_{\text{I}}    &=& \frac{2}{3} \kappa p^2_{\phi}, \\
\Omega_{\text{II}}   &=& \frac{4}{3} \Lambda,  \\
\Omega_{\text{III}}  &=& \frac{1}{9} \kappa^2 \gamma^2 \xi^2  p^4_{\phi}, \\
\Omega_{\text{IV}}   &=& \frac{4}{9} \kappa \Lambda p^2_{\phi} \gamma^2 \xi^2, \\
\Omega_{\text{V}}    &=& \frac{4}{9} \Lambda^2  \gamma^2 \xi^2.
\end{eqnarray}

Equation (\ref{mainequation}) is, in fact, a modified Friedmann equation 
\begin{equation}
H^2 = \frac{8\pi G_{\text{eff}}}{3} \rho +\frac{\Lambda_{\text{eff}}}{3},
\end{equation}
where the effective constants are expressed as follow
\begin{eqnarray}
 G_{\text{eff}}        &=& G\left[1-\frac{\rho}{\rho_{\text{c}}}  \right],     \\
\Lambda_{\text{eff}}   &=& 
\Lambda \left[1-2\frac{\rho}{\rho_{\text{c}}}-\frac{\Lambda}{\kappa
\rho_{\text{c}} } \right],
\end{eqnarray}
and 
\begin{eqnarray}
\rho &=&  \frac{p^2_{\phi}}{2{|p|}^3},      \\
\rho_{\text{c}} &=& \frac{3}{\kappa \gamma^2 \bar{\mu}^2 |p|}.
\end{eqnarray}

We will study the solutions of the equation (\ref{mainequation}) for both
$\bar{\mu}-$scheme and $\mu_0-$scheme.

The organisation of the text is the following. In section \ref{sec:NoLambda}
 we consider models with  $\Lambda=0$. We find solutions  of the equations 
(\ref{mainequation})  for both $\bar{\mu}-$scheme and $\mu_0-$scheme.
Next, in section \ref{sec:Lambda} we add to our considerations a non-vanishing
cosmological constant $\Lambda$. We carry out an analysis similar to the
case without lambda. We find analytical solutions of the
equation(\ref{mainequation}) for $\bar{\mu}-$scheme.
Then, we study the behaviour of this case in $\mu_0-$scheme.
In section \ref{sec:summary} we summarise the results.

\section{Models with $\Lambda=0$} \label{sec:NoLambda}

In this section we begin our considerations with the model without $\Lambda$. 
Equation (\ref{mainequation}) is then simplified to the form
\begin{eqnarray}
\left(\frac{dp}{dt} \right)^2 = \Omega_{\text{I}} {|p|}^{-1}-\Omega_{\text{III}}{|p|}^{2n-3}. 
\label{equation1}
\end{eqnarray}
We solve this equation for both $\bar{\mu}-$scheme and $\mu_0-$scheme in the present section.

\subsection{$\bar{\mu}-$scheme \ $(n=-1/2)$ }

In the $\bar{\mu}-$scheme, as is explained in the Appendix \ref{Appendix1}, the $\bar{\mu}$ is
expressed as 
\begin{equation}
\bar{\mu} = \sqrt{\frac{\Delta}{|p|}},
\end{equation}
where $\Delta \equiv 2 \sqrt{3} \pi \gamma l^2_{\text{Pl}}$ is the area gap.
So $\xi=\sqrt{\Delta}$ and $n=-1/2$.

\begin{figure}[ht!]
\centering
\includegraphics[width=6cm,angle=270]{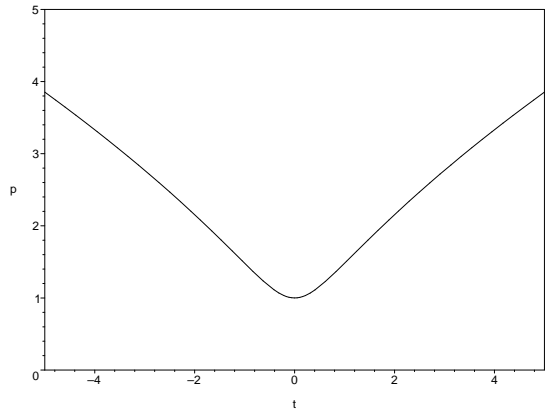}
\caption{Typical solution with $\Lambda=0$ in the $\bar{\mu}-$scheme.
The canonical variable $p$ is expressed in the $[l^2_{\text{Pl}}]$ units and time $t$ in the $[l_{\text{Pl}}]$ units.}
\label{solution1}
\end{figure}

To solve (\ref{equation1}) in the considered scheme we introduce a new
dependent variables $u$ in the form  
\begin{equation}
|p(t)|=u^{1/3}(t).
\end{equation}
With use of the variable $u$ the equation (\ref{equation1}) takes the form
\begin{eqnarray}
\left(\frac{du}{dt} \right)^2 = 9\Omega_{\text{I}} u(t)-9\Omega_{\text{III}}. 
\end{eqnarray}
and has a solution in the form of a second order polynomial  
\begin{eqnarray}
u(t) = \frac{\Omega_{\text{III}} }{\Omega_{\text{I}}} +\frac{9}{4} \Omega_{\text{I}} t^2 -
\frac{18}{4} \Omega_{\text{I}} C_1 t +\frac{9}{4} \Omega_{\text{I}} C_1^2,
\end{eqnarray}
where $C_1$ is a constant of integration. We can choose now $ C_1=0$, so that
the minimum of $u(t)$ occurs for $t=0$. Going back to the canonical variable
$p$ we obtain a bouncing solution
\begin{equation}
p(t) = \text{sgn}(p)   \left[ \frac{\Omega_{\text{III}} }{\Omega_{\text{I}}} +\frac{9}{4} \Omega_{\text{I}} t^2  \right]^{1/3}.
\label{sol1}
\end{equation}
The main property of this solutions is that $p(t)$ never reaches zero value for non vanishing $p_{\phi}$.
The minimal value of $p(t)$ is given by 
\begin{equation}
|p_{\text{min}}| =\gamma l_{\text{Pl}}^2 \left[ \frac{8\sqrt{3}}{3} \pi^2  \left(\frac{p_{\phi}}{ l_{\text{Pl}}} \right)^2 \right]^{1/3}.
\label{pmin}
\end{equation}
We show the solution (\ref{sol1}) in the Fig. \ref{solution1}. 
The dynamical behaviour in this model is simple. For negative times we have a
contracting pre-Big Bang 
Universe. For $t>0$ we have a Big Bang evolution from minimal $p_{\text{min}}$
(\ref{pmin}). It is, however, convenient to call this kind of stage Big Bounce
rather than Big Bang because of initial singularity avoidance.    

\subsection{$\mu_0-$Scheme \ $(n=0)$ } 
\begin{figure}[ht!]
\centering
\includegraphics[width=6cm,angle=270]{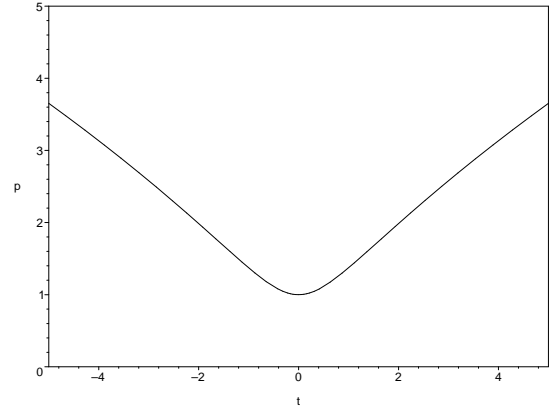}
\caption{Typical solution with $\Lambda=0$ in the $\mu_0-$scheme. The canonical variable $p$ is expressed in $[l^2_{\text{Pl}}]$ units and time $t$ in the $[l_{\text{Pl}}]$ units.}
\label{solution2}
\end{figure}

In the $\mu_0-$Scheme the $\bar{\mu}$ is
expressed as 
\begin{equation}
\bar{\mu}=\mu_0 = \frac{3\sqrt{3}}{2}.
\end{equation}
So $\xi=\mu_0$ and $n=0$.
To solve (\ref{equation1}) in this scheme, we change the time  
\begin{equation}
dt = \sqrt{|p(u)|} du 
\end{equation}
and introduce a new variable $y$ as follows
\begin{equation}
y(u) = |{p}(u)|^2. 
\end{equation}
Applying this new parametrisation to the equation (\ref{equation1}) leads to
the equation in the form
\begin{equation}
\left(\frac{dy}{du} \right)^2 = 4\Omega_{\text{I}} y(u)-4\Omega_{\text{III}} 
\end{equation}
which has a solution 
\begin{eqnarray}
y(u) = \frac{\Omega_{\text{III}} }{\Omega_{\text{I}}} + \Omega_{\text{I}} u^2 -
2 \Omega_{\text{I}} C_1 u + \Omega_{\text{I}} C_1^2. 
\end{eqnarray}
We can now choose $C_1=0$ so that the minimal value of
$y(u)$ occurs for $u=0$. Now we can 
go back to the initial parameters $p$ and $t$, then  
\begin{eqnarray}
p(u) &=& \text{sgn}(p)   \sqrt{ \frac{\Omega_{\text{III}} }{\Omega_{\text{I}}} + \Omega_{\text{I}} u^2  } \\
t(u) &=& \int_0^u du' \left[  \frac{\Omega_{\text{III}} }{\Omega_{\text{I}}} + \Omega_{\text{I}} u'^{2} \right]^{1/4}  \label{time1}
\end{eqnarray}
Introducing the variable 
\begin{equation}
x = \frac{\Omega_{\text{I}} }{\sqrt{\Omega_{\text{III}}}} u,
\end{equation}
we can rewrite integral (\ref{time1}) to the simplest form 
\begin{eqnarray}
t(u)  &=&  \frac{\sqrt{\Omega_{\text{III}}}} {\Omega_{\text{I}}} \left(
\frac{\Omega_{\text{III}} }{\Omega_{\text{I}}}   \right)^{1/4}
\int_0^{\frac{\Omega_{\text{I}} }{\sqrt{\Omega_{\text{III}}}} u } dx
\left[1+x^2 \right]^{1/4},
 \end{eqnarray}
and the solution of such integral is given by 
\begin{eqnarray}
\int_0^x \left[1+y^2 \right]^{1/4} dy  &=&  \frac{2}{3} x\left(1+x^2 \right)^{1/4}  \nonumber \\ 
   &+& \frac{1}{3} x {_2F_1}\left[ \frac{1}{2},\frac{3}{4}, \frac{3}{2} ; -x^2
   \right],
 \end{eqnarray}
where ${_2F_1}$ is the hypergeometric function, defined as
\begin{equation}
_pF_q[a_1,...,a_p ; b_1, ... ,b_q ; x] = \sum_{k=0}^{\infty}
\frac{(a_1)_k...(a_p)_k}{(b_1)_k...(b_q)_k}\frac{x^k}{k!},
\end{equation}
where $(a)_k$ is the Pochhammer symbol defined as follow
\begin{equation}
(a)_k \equiv \frac{\Gamma (a+k)}{\Gamma (a)} = a (a+1) \dots (a+k-1) .
\end{equation}
This solution is very similar to the one in $\bar{\mu}-$scheme. However, the time parametrisation 
is expressed in a more complex way. We show this solution in the Fig. \ref{solution2}.
In this case minimal value of $p(t)$ is expressed as
\begin{equation}
|p_{\text{min}}| = \frac{3}{2}\gamma  l_{\text{Pl}}^2  \sqrt{ \frac{\pi}{2}  \left(\frac{p_{\phi}}{ l_{\text{Pl}}} \right)^2 }.
\label{pmin2}
\end{equation}

\section{Models with $\Lambda \neq 0$} \label{sec:Lambda}

In this section, we investigate the general model with non vanishing cosmological constant.
It will be useful to write equation (\ref{mainequation}) in the 
form 
\begin{eqnarray}
\left(\frac{dp}{dt} \right)^2 &=& \left[ \Omega_{\text{I}} {|p|}^{-1}+   \Omega_{\text{II}}{|p|}^2 \right]\times    \nonumber \\
                              &\times& \left[ 1- \frac{\Omega_{\text{III}}}{\Omega_{\text{I}}}{|p|}^{2n-2} - 
            \frac{\Omega_{\text{V}}}{\Omega_{\text{II}}}{|p|}^{2n+1}
            \right].
\label{equation22}
\end{eqnarray}
We see that when we perform the multiplication in this equation and define
\begin{equation}
\Omega_{\text{IV}} = \frac{ \Omega_{\text{II}} \Omega_{\text{III}} }{
\Omega_{\text{I}} }+\frac{\Omega_{\text{I}}\Omega_{\text{V}}
}{\Omega_{\text{II}}},
\end{equation}
we recover equation (\ref{mainequation}). In this and the next section we use equivalently $\Lambda_0=\frac{3}{\xi^2\gamma^2}$ and
$\alpha=\frac{\kappa}{6}p^2_{\phi}$ to simplify notation. 

\subsection{$\bar{\mu}-$scheme \ $(n=-1/2)$ }
In this case equation (\ref{equation22}) can be rewritten to the form 
\begin{equation}
\left(\frac{dp}{dt} \right)^2 =\gamma_{\text{I}} {|p|}^2 +\gamma_{\text{II}}
{|p|}^{-1}-\gamma_{\text{III}}{|p|}^{-4},
\label{equation2}
\end{equation}
where the parameters are defined as
\begin{eqnarray}
\gamma_{\text{I}}   &=&  \Omega_{\text{II}} - \Omega_{\text{V}} = 
\frac{4}{3}\Lambda \left[1-\frac{\Lambda}{3}\gamma^2 \xi^2 \right],   \\
\gamma_{\text{II}}  &=&  \Omega_{\text{I}} - \Omega_{\text{IV}} = 
 \frac{2}{3}\kappa p^2_{\phi}\left[1-2\frac{\Lambda}{3}\gamma^2 \xi^2 \right] ,   \\
\gamma_{\text{III}} &=&  \Omega_{\text{III}}=  \frac{1}{9} \kappa^2 \gamma^2 \xi^2  p^4_{\phi} .      
\end{eqnarray}
To solve equation (\ref{equation2}) we re-parametrise the time variable
\begin{equation}
dt = |p(u)|^3du, 
\end{equation}
and introduce a new variable $p$ as follows
\begin{equation}
y(u) = |p(u)|^3. 
\end{equation}
This change of variables leads to the equation in the form
\begin{equation}
\left(\frac{dy}{du} \right)^2 =9\gamma_{\text{I}} y^2 (y-y_1)(y-y_2),
\label{equation3}
\end{equation}
where 
\begin{eqnarray}
y_1 &=& \frac{1}{6} \frac{\kappa p^2_{\phi} \gamma^2 \xi^2 }{\left[
1-\frac{\Lambda}{3}\gamma^2 \xi^2 \right] },        \\
y_2 &=& -\frac{1}{2} \frac{\kappa  p^2_{\phi}}{\Lambda}.
\end{eqnarray}

There are three general types of solutions corresponding to the values
of the parameters $(\gamma_{\text{I}},y_1,y_2)$.
We summarise these possibilities in the table below.

\begin{equation}
\begin{array}{|c||c|c|c|}
\hline                              & 1  & 2  & 3  \\
\hline \hline
\Lambda                             & <0 & >0 & >0 \\
  1-\frac{\Lambda}{3}\gamma^2 \xi^2 & >0 & >0 & <0 \\
\hline \hline
\gamma_{\text{I}}                   & <0 & >0 & <0 \\
y_1                                 & >0 & >0 & <0 \\
y_2                                 & >0 & <0 & <0 \\
\hline
\end{array} \nonumber
\end{equation}

\begin{figure}[ht!]
\centering
\includegraphics[width=6cm,angle=270]{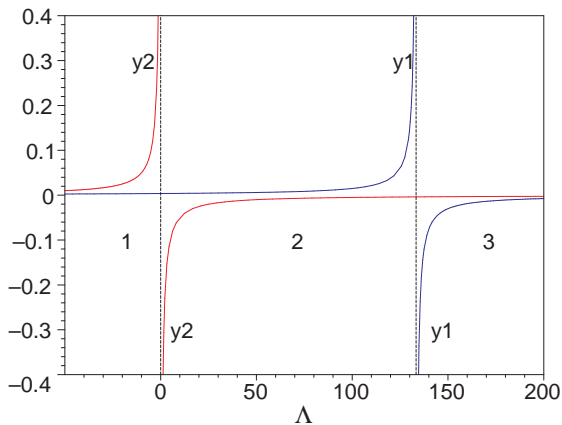}
\caption{Parameters $y_1$ and $y_2$ as functions of $\Lambda$. Region 1
corresponds to
oscillatory solution, region 2 to bouncing solution. There is no physical 
solutions in region 3. The parameter $\Lambda$ is expressed in units of $[m^2_{\text{Pl}}]$.}
\label{roots1}
\end{figure}

It is important to notice that the product 
$ \Upsilon  \equiv \gamma_{\text{I}} \cdot y_1 \cdot  y_2 = -12 \frac{\alpha^2}{\Lambda_{0}} $ is negative in all 
cases. This property will be useful to solve equation (\ref{equation3}).
In the Fig. \ref{roots1} we show values of the roots $y_1$ and $y_2$ as
functions of $\Lambda$. 
Thus, there are two values 
of cosmological constants where the signs of the roots change, namely $\Lambda=0$
and
\begin{equation}
\Lambda_{0,\bar{\mu}} = \frac{\sqrt{3} m^2_{\text{Pl}}}{2\pi \gamma^3}  = 133.4 m^2_{\text{Pl}},
\end{equation} 
where we have used the value of the $\gamma=\ln 2/(\pi \sqrt{3})$  calculated
in the work \cite{Ashtekar:1997yu}.
More recent investigation of the black hole entropy indicate however that 
the value of the Barbero-Immirzi parameter is $\ln 2 /\pi \leq \gamma \leq \ln
3/\pi$ \cite{Domagala:2004jt}.
In particular, Meissner has calculated $\gamma_{M}=0.12738$ \cite{Meissner:2004ju}.  
However, this freedom in the choice of $\gamma$ does not change the qualitative results.
Only the region of the parameter space where the particular kind of motion is
allowed can be shifted.  

We now perform a change of variables in equation (\ref{equation3})  in the form
\begin{equation}
w(u)  = \frac{1}{y(u)},
\end{equation}
and 
\begin{equation}
\chi = 3 \sqrt{|\Upsilon|} u.
\end{equation}
We also introduce the parameters 
\begin{eqnarray}
w_1 = \frac{1}{y_1} , \ \  w_2 = \frac{1}{y_2},
\end{eqnarray}
and then equation (\ref{equation3}) takes the form 
\begin{equation}
\left(\frac{dw}{d\chi} \right)^2 =  - (w_1-w)(w_2-w).
\label{equation4}
\end{equation}
The equation (\ref{equation4}) is the equation of shifted harmonic 
oscillator and its solution is
\begin{equation}
w(\chi) = \frac{\Lambda_{0,\bar{\mu}}}{6\alpha} \left[ 1-2\frac{\Lambda}{\Lambda_{0,\bar{\mu}}}  + \cos \left(\chi\right) \right].
\end{equation}
where we set the integration constant to zero.
It is now possible to
go back to the initial parameters $p$ and $t$ which are expressed as follows
\begin{eqnarray}
p(u) &=&  \text{sgn}(p)  \frac{ \sqrt[3]{ \frac{6\alpha}{\Lambda_{0,\bar{\mu}}}
} }{ \sqrt[3]{ 1-2\frac{\Lambda}{\Lambda_{0,\bar{\mu}}} +  \cos
\left(3\sqrt{|\Upsilon|}u\right) }},   \label{sol11}    \\
t(u) &=& \int_0^u du^{'} p^3(u^{'}),  \label{sol22}
\end{eqnarray}
The integral (\ref{sol22}) can be easily solved but solutions depend on the parameters in (\ref{sol11}). 

We define now an integral 
\begin{equation}
I(x,a) = \int_0^x \frac{dx'}{a+\cos x'}
\label{mainintegral1}
\end{equation}
so the expression (\ref{sol22}) can now be written as
\begin{equation}
t(u) = \frac{1}{3\sqrt{|\Upsilon|}}  I\left( 3\sqrt{|\Upsilon|}u, 1-2\frac{\Lambda}{\Lambda_{0,\bar{\mu}}}   \right) 
\end{equation}
Solutions of integral (\ref{mainintegral1}) depend on the value of parameter
$a$ and read
\begin{eqnarray}
I(x,|a|>1)   &=&   \frac{2}{\sqrt{a^2-1}} \arctan 
\left[ \frac{(a-1)\tan\left( \frac{x}{2} \right)}{\sqrt{a^2-1}} \right], \nonumber \\
I(x,a= 1) &=&  \tan \left(\frac{x}{2}  \right), \nonumber   \\
I(x,0<|a|<1) &=&   -\frac{2}{\sqrt{1-a^2}} \text{arctanh}
 \left[ \frac{(a-1)\tan\left( \frac{x}{2} \right)}{\sqrt{1-a^2}} \right], \nonumber \\
I(x,a=0) &=& \ln \left( \frac{1}{\cos x}  - \tan x  \right). \nonumber
\end{eqnarray} 

\begin{figure}[ht!]
\centering
\includegraphics[width=6cm,angle=270]{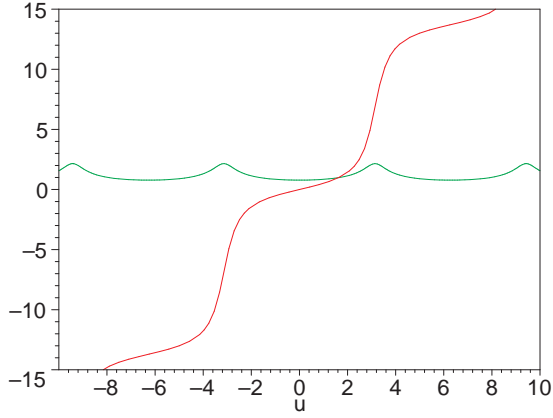}
\caption{Parametric solution with $\Lambda<0$. Oscillating curve (green) is
$|p(u)|$ and the increasing curve (red) is $t(u)$. The canonical variable $p$ is
 expressed in the $[l^2_{\text{Pl}}]$ units and time $t$ in the $[l_{\text{Pl}}]$ units.}
\label{solution4}
\end{figure}

In Fig. \ref{solution4} we show the solution for $\Lambda<0$ -- or equivalently with parameter $a>1$. 

\begin{figure}[ht!]
\centering
\includegraphics[width=6cm,angle=270]{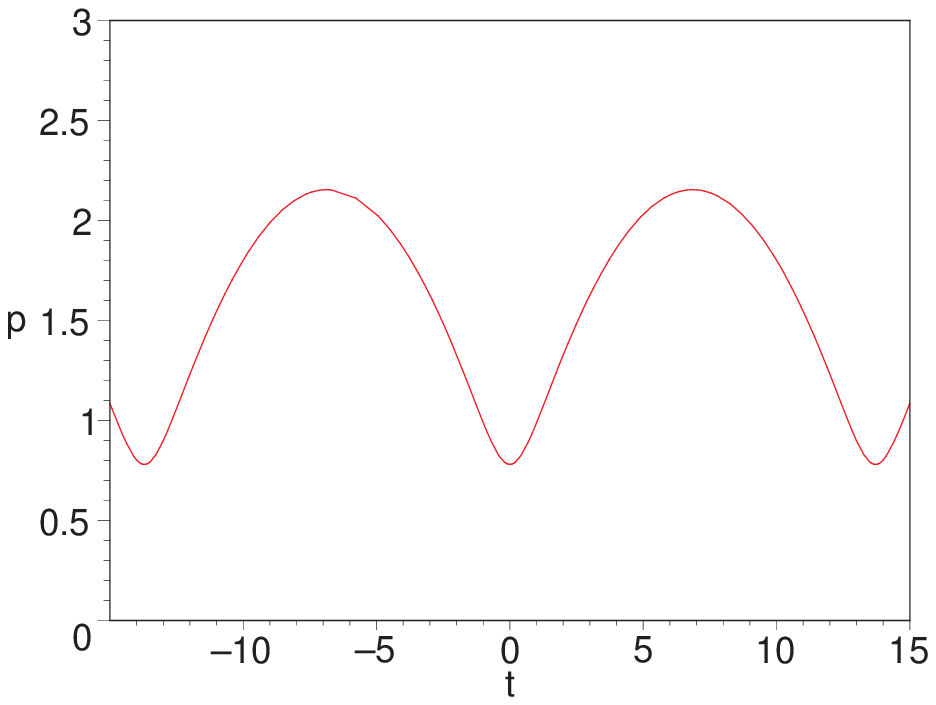}
\caption{Solution $|p(t)|$with $\Lambda<0$. The canonical variable $p$ is expressed in
 units of $[l^2_{\text{Pl}}]$ and time $t$ in $[l_{\text{Pl}}]$.}
\label{solution5}
\end{figure}

In Fig. \ref{solution5} we show the solution for
$0<\Lambda<\Lambda_{0,\bar{\mu}}$ which corresponds to $0<|a|<1$. 
\begin{figure}[ht!]
\centering
\includegraphics[width=6cm,angle=270]{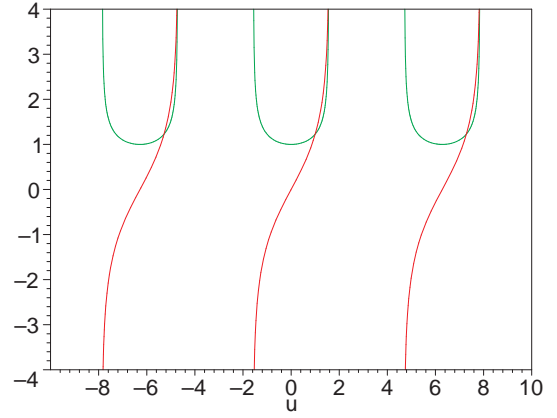}
\caption{Parametric solution with $0<\Lambda<\Lambda_{0,\bar{\mu}}$. Top curve
(green) shows $|p(u)|$ and bottom curve (red) is $t(u)$. The canonical variable $p$ is
 expressed in units of $[l^2_{\text{Pl}}]$ and time $t$ in $[l_{\text{Pl}}]$.}
\label{solution6}
\end{figure}

\begin{figure}[ht!]
\centering
\includegraphics[width=6cm,angle=270]{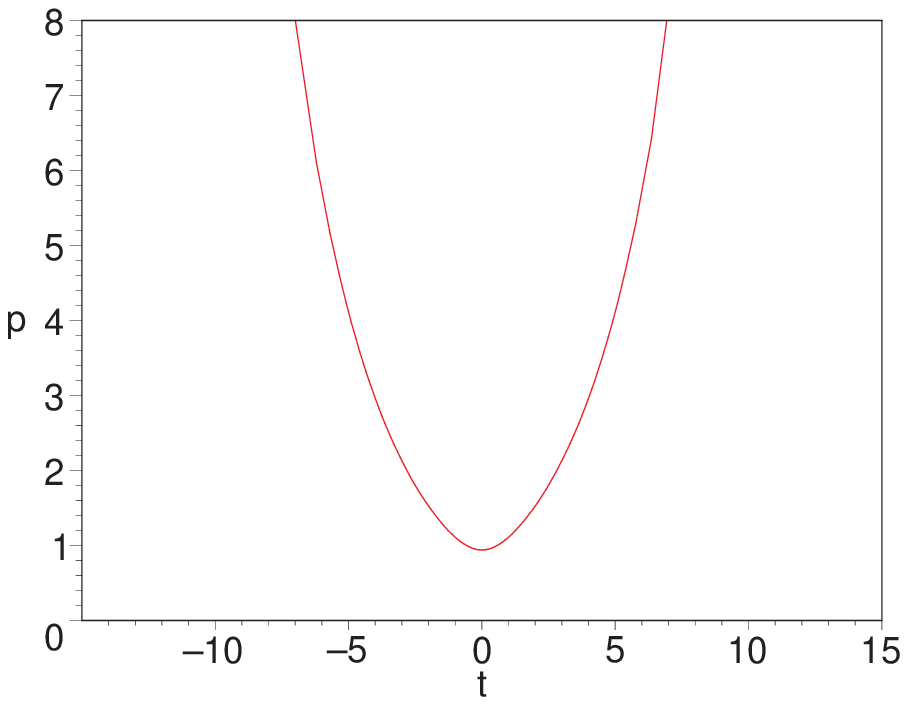}
\caption{Solution $|p(t)|$ with $0<\Lambda<\Lambda_{0,\bar{\mu}}$. The canonical variable $p$
 is expressed in units of $[l^2_{\text{Pl}}]$ and time $t$ in $[l_{\text{Pl}}]$.}
\label{solution7}
\end{figure}

\subsection{$\mu_0-$Scheme \ $(n=0)$ }

In this section we study the last case in which $n=0$ and $\xi = 3\sqrt{3}/2$. 
Equation (\ref{equation22}) can then be written in the form 
\begin{equation}
\left(\frac{dp}{dt} \right)^2 = {|p|}^{-3} \left( \Omega_{\text{I}} +
\Omega_{\text{II}} {|p|}^3 \right) W(|p|),
\label{equation6}
\end{equation}
where
\begin{equation}
W(|p|)=  {|p|}^3 a_3 + {|p|}^2a_2 +{|p|} a_1 +a_0,
\label{polynomial}
\end{equation}
and the polynomial's coefficients are  expressed as
\begin{eqnarray}
a_0 &=&  -\frac{1}{6} \kappa p^2_{\phi} \gamma^2 \xi^2  =-\frac{3\alpha}{\Lambda_{0,\mu_0}},\\
a_1 &=&  0,\\
a_2 &=&  1,\\
a_3 &=&  -\frac{\Lambda}{3} \gamma^2 \xi^2  = -\frac{\Lambda}{\Lambda_{0,\mu_0}}.
\end{eqnarray}
The discriminant of polynomial (\ref{polynomial}) is
\begin{eqnarray}
\tilde{\Delta} = \frac{ a_2^2 a_1^2 - 4 a_1^3 a_3 -4a_2^3a_0+18a_0a_1a_2a_3-27a_0^2a_3^2}{a_3^4}.
\end{eqnarray}
Inserting the values of parameters $\{a_0,a_1,a_2,a_3\}$ listed above we obtain
\begin{eqnarray}
\tilde{\Delta} =\left( \frac{\Lambda}{\Lambda_{0,\mu_0}}\right)^{-4} \left[\frac{12\alpha}{\Lambda_{0,\mu_0}} 
-27 \left( \frac{3\alpha}{\Lambda_{0,\mu_0}} \frac{\Lambda}{\Lambda_{0,\mu_0}} \right)^2 \right].
\end{eqnarray}
For $\tilde{\Delta}>0$, or
\begin{equation}
\frac{4}{81} >   \left( \frac{\alpha}{\Lambda_{0,\mu_0}} \right) \left( \frac{\Lambda}{\Lambda_{0,\mu_0}} \right)^2
\label{rel}
\end{equation}
polynomial (\ref{polynomial}) has three real roots. 
When relation (\ref{rel}) is fulfilled and $\Lambda>0$, oscillatory 
solutions occur. For $\Lambda<0$ equation (\ref{equation6}) has bouncing type solutions.
This can be seen when we redefine equation (\ref{equation6}) to the point particle form.
This approach is useful for qualitative analysis and will be fully used in the next section.
However, here we will use it to distinguish between different types of solutions.

Introducing a new time variable 
\begin{equation}
dt =\frac{ {|p|}^{3/2}}{ \sqrt{ \Omega_{\text{I}} +\Omega_{\text{II}} {|p|}^3
} } du,
\end{equation}
we can rewrite equation  (\ref{equation6}) to the form 
\begin{equation}
\mathcal{H}=\frac{1}{2}p'^{2}+V(p)=E=-\frac{3}{2}\frac{\alpha}{\Lambda_{0,\mu_0}},
\label{HamConstr}
\end{equation}
where the potential function
\begin{equation}
V(p)=-\frac{1}{2}|p|^{2}(1- \frac{\Lambda}{\Lambda_{0,\mu_0}} |p|).
\label{pot}
\end{equation}
Equation (\ref{HamConstr}) has the form of Hamiltonian constraint for a
point particle in a potential well. We see that for $\Lambda<0$ potential (\ref{pot})
has only one extremum (for $p=0$) and only a bouncing solution is possible.
For $\Lambda<0$ potential (\ref{pot}) has a minimum for $|p_{\text{min}}|=\frac{2}{3} \frac{\Lambda_{0,\mu_0}}{\Lambda}$.
In this case physical solutions correspond to the condition
$E>V(|p_{\text{min}}|)$ and the energy of the 
imagined particle in the potential well is greater than the minimum of well.
The particle then oscillates between the boundaries of the potential.
The condition  $E>V(|p_{\text{min}}|)$ is equivalent to relation (\ref{rel})
calculated from the discriminant. 

Upon introducing the parameters
\begin{eqnarray}
g_2 &:=&  \frac{1}{12} a_2^2 -\frac{1}{4} a_3 a_1, \\
g_3 &:=&  \frac{1}{48} a_3a_2a_1-\frac{1}{216} a_2^3 - \frac{1}{16} a_3^2 a_0, 
\end{eqnarray}
and the variable $v$
\begin{equation}
|p| = \frac{4}{a_3} v - \frac{a_2}{3a_3}
\end{equation}
then equation (\ref{equation6}) takes on the form of the Weierstrass equation
\begin{equation}
\left( \frac{dv}{du} \right)^2 = 4 v^3 - g_2v-g_3.
\label{Weierstrass}
\end{equation}
Detailed analysis and plots of solutions of this equation for 
different values of parameters can be found in the appendix to the article \cite{Dabrowski:2004hx}.
The solution of equation (\ref{Weierstrass})  is the Weierstrass $\wp$-function
\begin{equation}
v(u) = \wp (u-u_0 ; g_2 , g_3 ),
\end{equation}
where 
\begin{eqnarray}
g_2 &=&  \frac{1}{12}, \\
g_3 &=&  -\frac{1}{216} +\frac{1}{16} \left( \frac{\Lambda}{\Lambda_{0,\mu_0}} \right)^2\frac{3\alpha}{\Lambda_{0,\mu_0}}.
\end{eqnarray}

So the parametric solution for the parameter $p$ is
\begin{equation}
p(u) = \text{sgn}(p) \frac{\Lambda_{0,\mu_0}}{\Lambda}\left[\frac{1}{3}-4\wp (u-u_0 ; g_2 , g_3 )  \right]
\end{equation}
The time variable can be expressed as the integral
\begin{equation}
t(u) = \int_0^{u}  du' \frac{\left(   \frac{4}{a_3} v(u') - \frac{a_2}{3a_3}\right)^{3/2}}{ \sqrt{ 
 \Omega_{\text{I}}+\Omega_{\text{II}} \left(  \frac{4}{a_3} v(u') - \frac{a_2}{3a_3}  \right)^3   }  }. 
\end{equation}

\begin{figure}[ht!]
\centering
\includegraphics[width=6cm,angle=270]{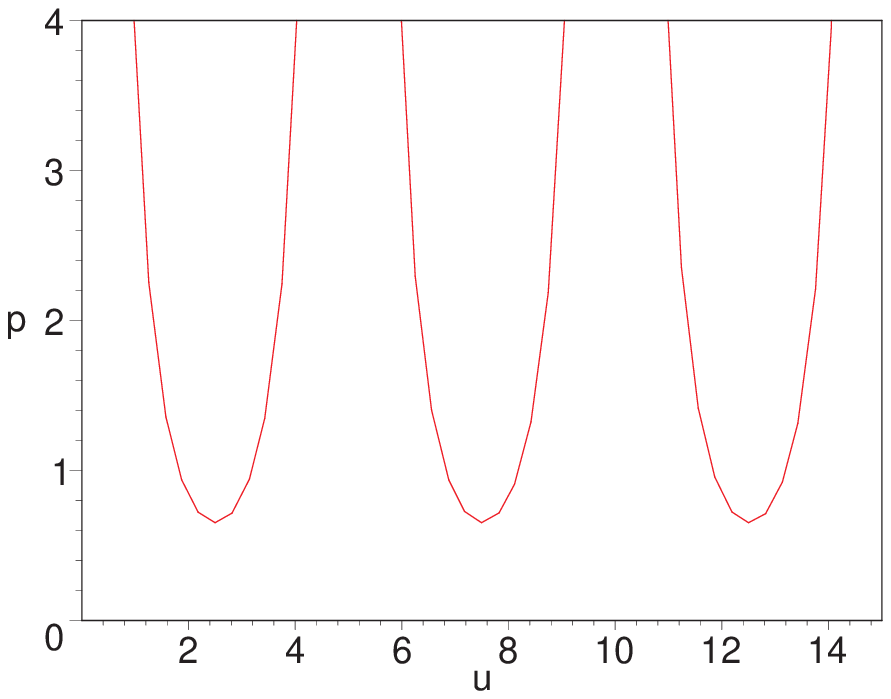}
\caption{Parametric solution $|p(u)|$ with $\Lambda<0$. The canonical variable $p$ is
 expressed in units of $[l^2_{\text{Pl}}]$, $u$ is dimensionless.}
\label{solution9}
\end{figure}

\begin{figure}[ht!]
\centering
\includegraphics[width=8cm,angle=0]{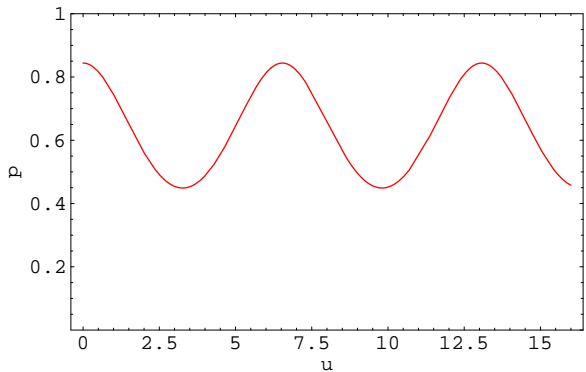}
\caption{Parametric solution $|p(u)|$ with $\Lambda>0$ and relation (\ref{rel})
satisfied. The canonical variable $p$ is
 expressed in units of $[l^2_{\text{Pl}}]$ and $u$ is dimensionless.}
\label{solution9a}
\end{figure}

In the Fig. \ref{solution9} and \ref{solution9a} we show an exemplary parametric
bounce solution with $\Lambda<0$ and a possible oscillatory solution with
$\Lambda>0$.

In general the solution is expressible as an explicit function of time, by
means of the so called Abelian functions. However, we chose not to employ them
here as the appropriate formulae are not as clear, and the commonly used
numerical packages do not allow for their direct plotting. The fact of 
existence of such solutions allows us to assume that the above integral is
well defined, and so is the solution itself.

\section{Qualitative methods of differential equations in study evolutional
paths}

The main advantage of using qualitative methods of differential equations (dynamical
systems methods) is the investigation of all solutions for all admissible initial
conditions. We demonstrate that dynamics of the model can be reduced to the form
of two dimensional autonomous dynamical system. In our case the phase space is
$2D$ $(p,\dot{p})$. First we can find the solutions corresponding to vanishing of
the right hand sides of the system which are called critical points. The
information about their stability and character is contained in the linearisation matrix around
a given critical point. In the considered case, the dynamical system is of the
Newtonian type. For such a system, the characteristic equation which determines
the eigenvalues of the linearisation matrix at the critical point is of the
form $\lambda^{2}+ \partial^{2} V(p)/\partial p^{2}$  where $V(p)$ is a
potential function and $|p|=a^{2}V_0^{2/3}$, $a$ being the scale factor. As it is well known for
dynamical systems of the Newtonian type, only two types of critical points are
admissible. If the diagram of the potential function is upper convex then
eigenvalues are real and of opposite signs, and the corresponding critical point is
of the saddle type. In the opposite case, if $\partial^{2} V(p)/\partial p^{2} > 0$
then the eigenvalues are purely imaginary and conjugate. The corresponding
critical point is of the centre type.

Equation (\ref{equation22}) can be written in the form
\begin{equation}
\frac{1}{4}\dot{p}^{2} = |p|^{2n-3}(\alpha+\frac{\Lambda}{3}|p|^{3})(|p|^{-2n+2}
-\gamma^{2}\xi^{2}(\alpha+\frac{\Lambda}{3}|p|^{3})),
\end{equation}
where $\alpha=\frac{\kappa}{6}p_{\phi}^{2}$ or equivalently as
\begin{equation}
\frac{1}{4}\frac{|p|^{-2n+3}}{\alpha+\frac{\Lambda}{3}|p|^{3}}\dot{p}^{2}=
p'^{2} = |p|^{2}(|p|^{-2n}-\gamma^{2}\xi^{2}\frac{\Lambda}{3}|p|) -
\gamma^{2}\xi^{2}\alpha,
\end{equation}
where we have made the following time reparametrisation $t \to u$
\begin{equation}
\frac{d}{d u} =
\frac{1}{2}\frac{|p|^{-n+3/2}}{\sqrt{\alpha+\frac{\Lambda}{3}|p|^{3}}}
\frac{d}{d t}.
\end{equation}

Now we are able to write the Hamiltonian constraint in the form analogous to the
particle of the unit mass moving in the one dimensional potential well
\begin{equation}
\mathcal{H}=\frac{1}{2}p'^{2}+V(p)=E=-\gamma^{2}\xi^{2}\frac{\alpha}{2},
\end{equation}
where the potential function
\begin{equation}
V(p)=-\frac{1}{2}|p|^{2}(|p|^{-2n}-\gamma^{2}\xi^{2}\frac{\Lambda}{3}|p|).
\end{equation}
As we can see, the constant $-\gamma^{2}\xi^{2}\alpha/2$ plays the role of the
total energy of
the fictitious particle. The domain admissible for motion in the configuration space
is determined by the condition $V(p)+\gamma^{2}\xi^{2}\alpha/2<0$.

A dynamical system of the Hamiltonian type has the following form
\begin{equation}
\begin{array}{l}
p'=y, \\
y'=-\frac{\partial V}{\partial p},
\end{array}
\end{equation}
where the prime denotes differentiation with respect to a new re-parametrised
time variable which is a monotonous function of the original, cosmological time.

The structure of the phase plane is organised by the number and location of
critical points. In our case, critical points in the finite domain are located
only on the line $p'=y=0$ and the second coordinate is determined form the equation
$-\partial V/\partial p=0$ which is
\begin{equation}
p\big((1-n)|p|^{-2n}-\gamma^{2}\xi^{2}\frac{\Lambda}{2}|p|\big) =0.
\end{equation}
The number of critical points depends on the value of $\Lambda$ and
$n$. We can distinguish two cases:
\begin{itemize}
\item{for $-1/2<n\le0$:}
\subitem{$\Lambda<0$: $(|p|,p')=(0,0)$;}
\subitem{$\Lambda>0$: $(|p|,p')=(0,0)$ and \\
$(|p|,p')=(\sqrt[2n+1]{\frac{2(1-n)}{\gamma^{2}\xi^{2}\Lambda}},0)$;}
\item{for $n=-1/2$:}
\subitem{$\Lambda\ne \frac{3}{\gamma^{2}\xi^{2}}$: $(|p|,p')=(0,0)$;}
\subitem{$\Lambda=\frac{3}{\gamma^{2}\xi^{2}}$: degenerate $V(p)=0$.}
\end{itemize}

The full analysis of the behaviour of trajectories requires investigation also at the
infinity. To this aim we introduce radial coordinates on the
phase plane for compactification of the plane by adjoining the circle at infinity
$p=\frac{r}{1-r}\cos{\theta}$, $y=\frac{r}{1-r}\sin{\theta}$.

\begin{figure*}
a)\includegraphics[scale=0.75]{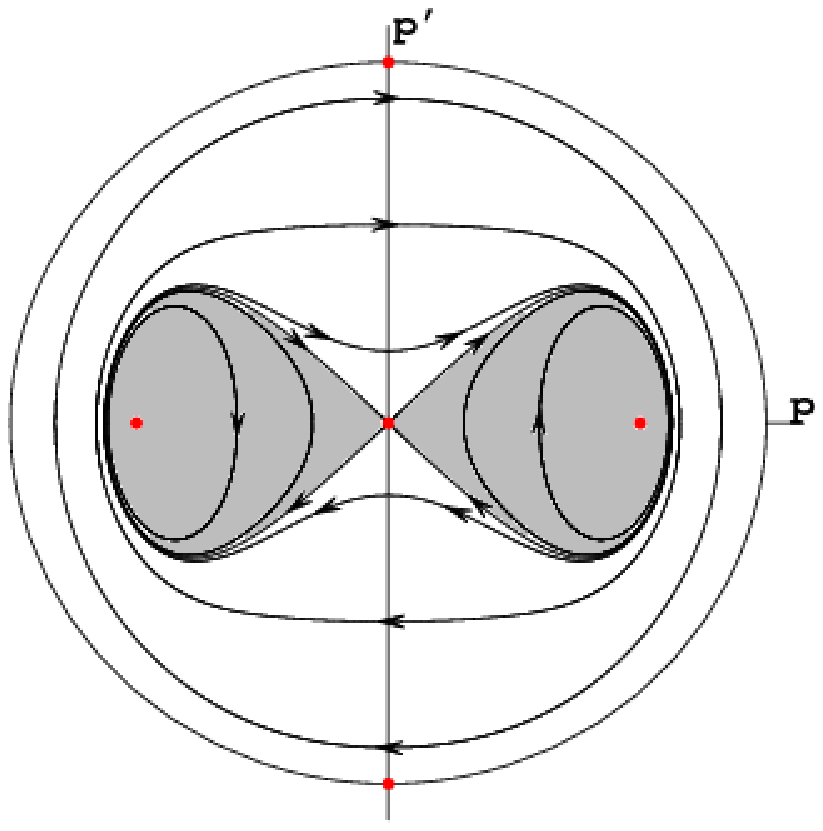}
\includegraphics[scale=1]{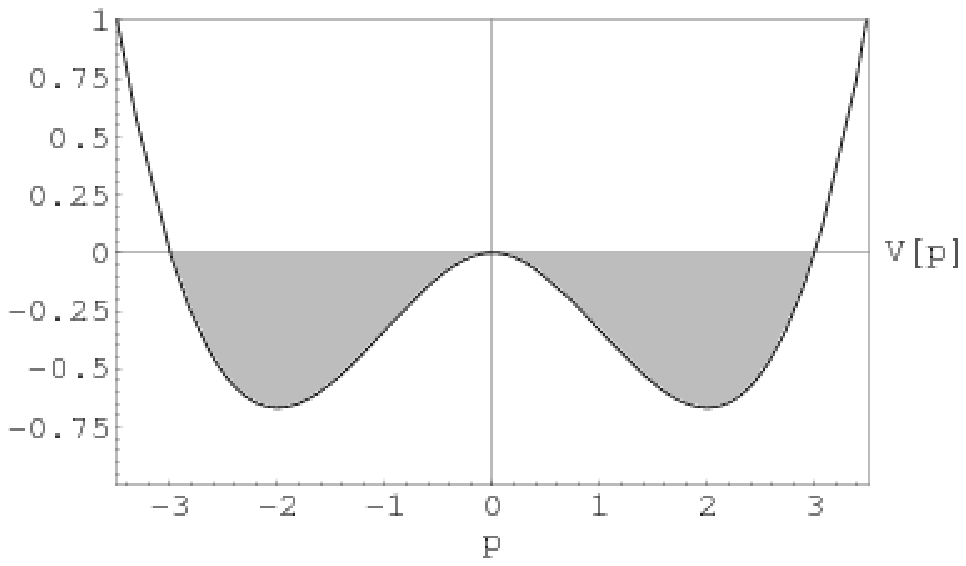}
b)\includegraphics[scale=0.75]{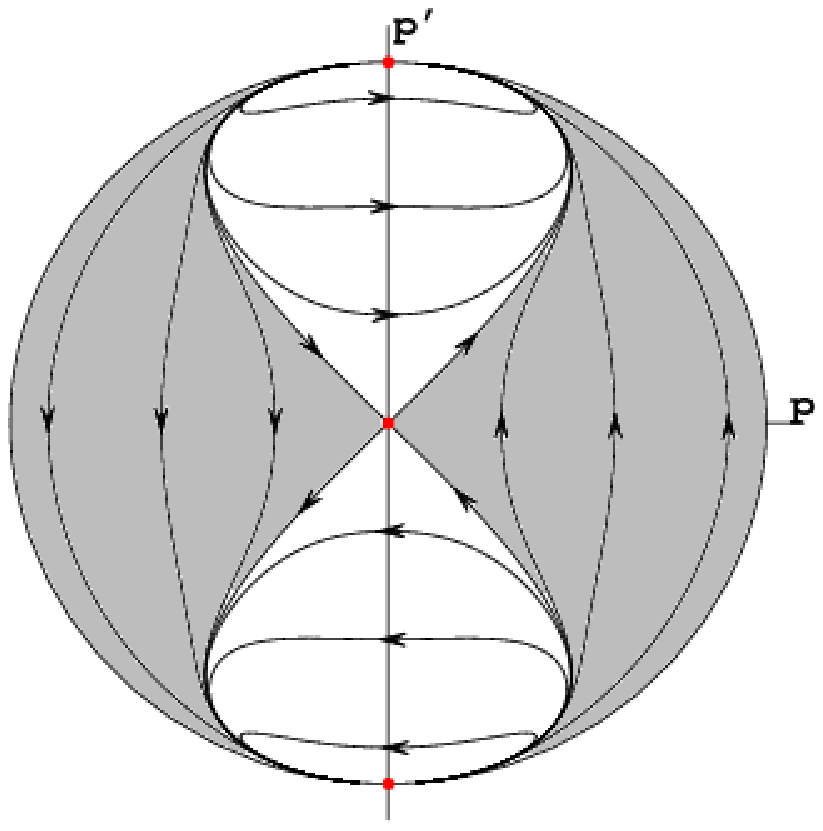}
\includegraphics[scale=1]{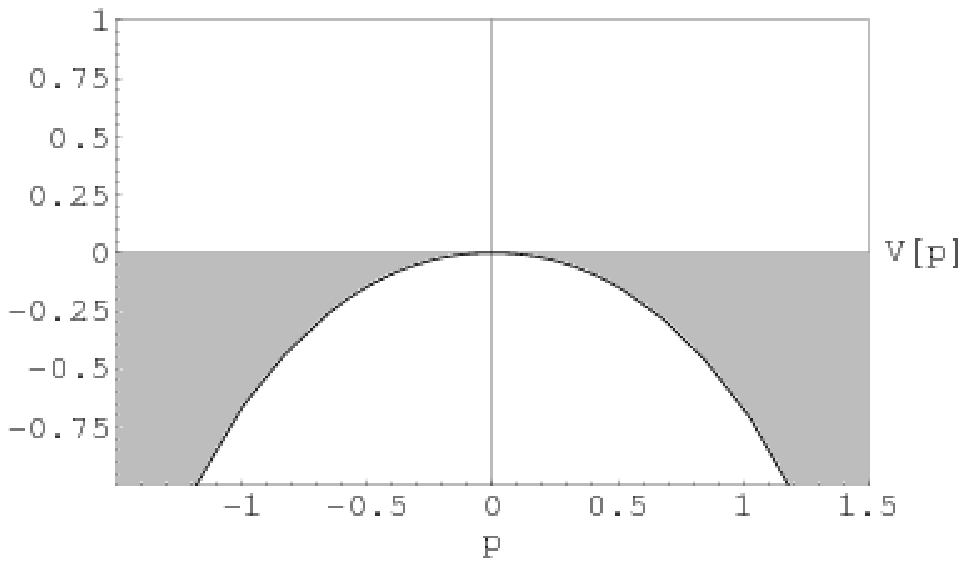}
\caption{The phase space diagram for the case $n=0$ and a) $\Lambda>0$ and b)
$\Lambda<0$. The physical domain admissible for motion is shaded.
Note that for the case (a) the boundary of admissible for motion is bounded 
by a homoclinic orbit and all solutions in this area are oscillating without 
initial and final singularities. All trajectories situated in physical 
region posses the minimum value of scale factor.}
\label{fig:1}
\end{figure*}

\begin{figure*}
a)\includegraphics[scale=0.75]{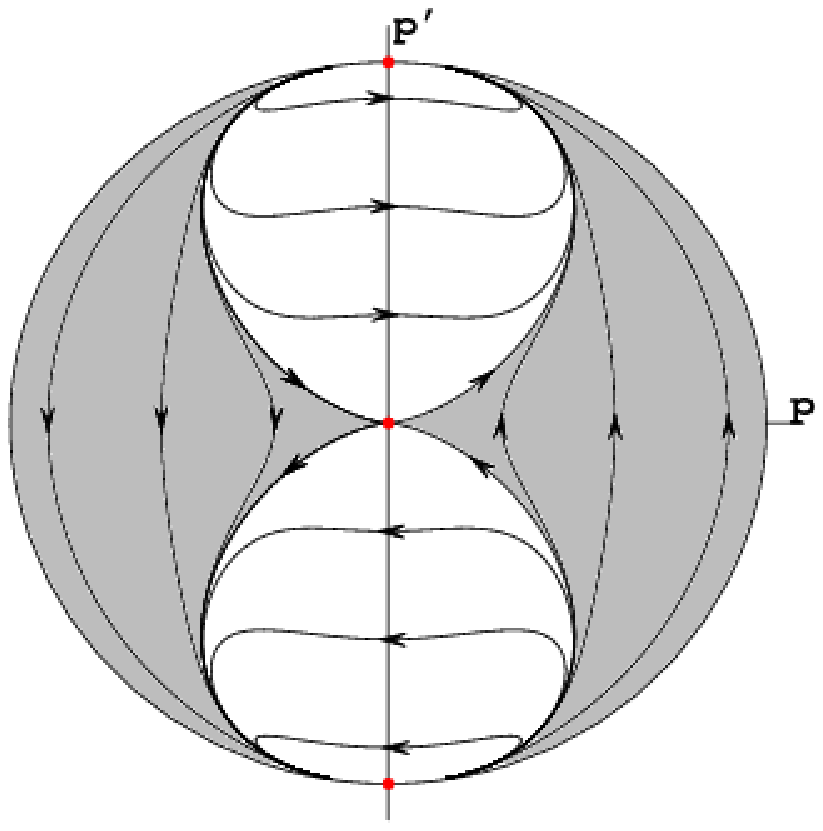}
\includegraphics[scale=1]{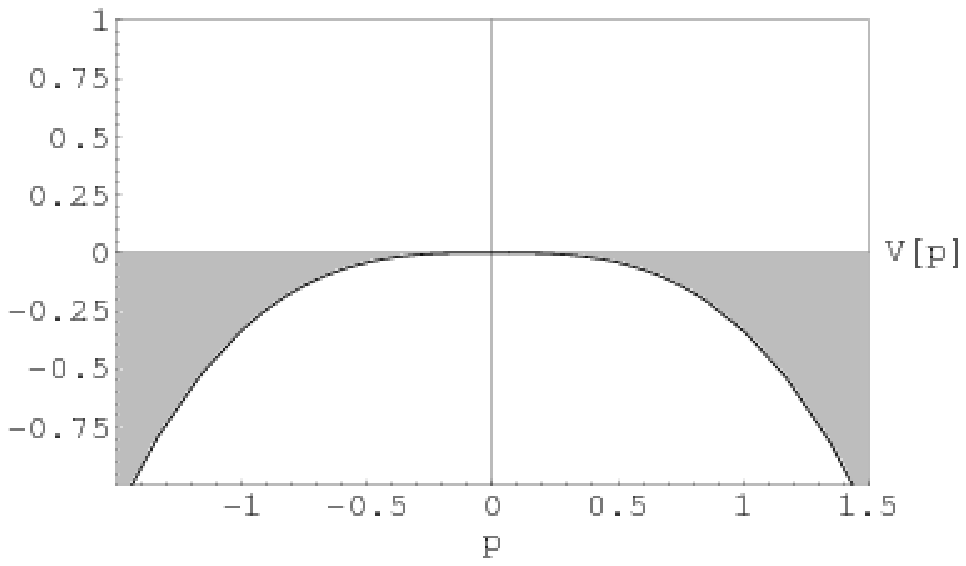}
b)\includegraphics[scale=0.75]{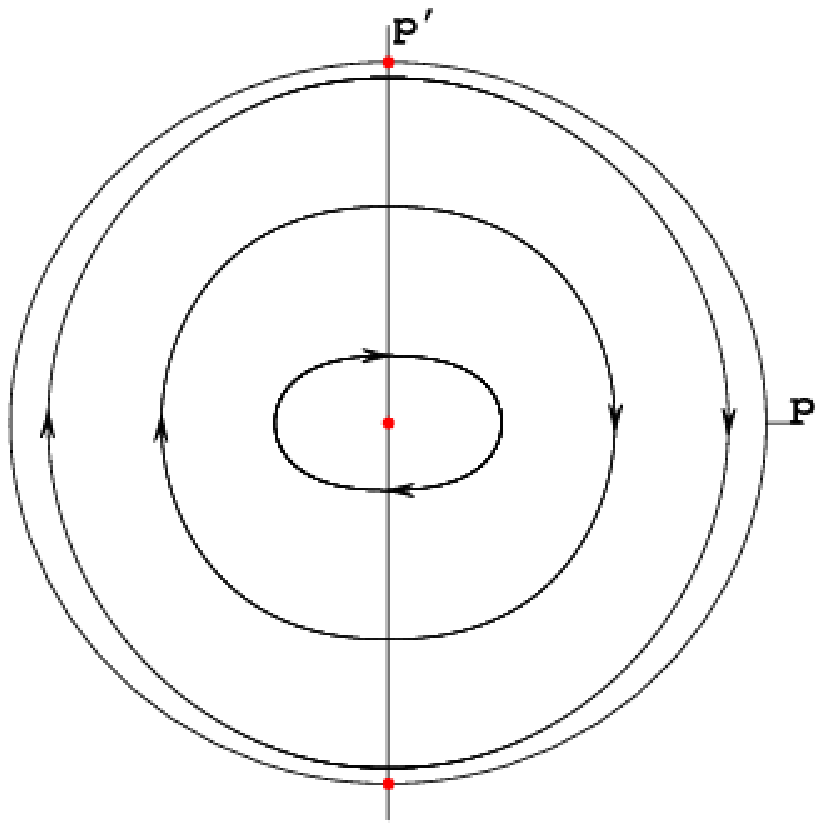}
\includegraphics[scale=1]{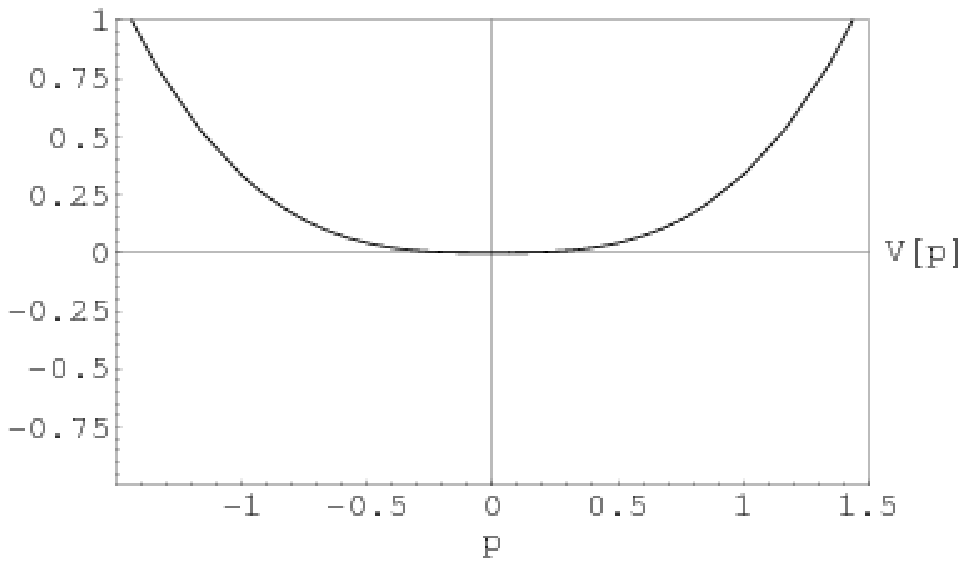}
\caption{The phase space diagram for the case $n=-1/2$ and a)
$\Lambda=\frac{1}{\gamma^{2}\xi^{2}} < \frac{3}{\gamma^{2}\xi^{2}}=\Lambda_0$ and b) $
\Lambda=\frac{5}{\gamma^{2}\xi^{2}} >\frac{3}{\gamma^{2}\xi^{2}}=\Lambda_0$. The physical
domain admissible for motion is shaded. In the second case there is no
physically allowed region for which $\alpha>0$.
This case is distinguished by behaviour at infinity when Hubble 
function is finite like for the de Sitter solution (see formula (\ref{deS})).
This state is the global attractor in the future. All solutions 
are of the bouncing type.  
}
\label{fig:2}
\end{figure*}

The phase portraits for both cases are shown at Figs. \ref{fig:1} and
\ref{fig:2}.

For the general case of $\Lambda>0$ we can write the parametric
equation of the boundary $\alpha=0$ of the physically admissible region in the
phase space, namely
\begin{equation}
\dot{p}^{2}=\frac{4}{3}\Lambda|p|^{2}(1-\gamma^{2}\xi^{2}\frac{\Lambda}{3}|p|^{2n+1}),
\end{equation}
where the dot denotes differentiation with respect to cosmological time $t$. This
equation greatly simplifies for the special case $n=-1/2$, and we receive the
value of the Hubble parameter at the boundary
\begin{equation}
\frac{1}{4}\frac{\dot{p}^{2}}{|p|^{2}} = H^{2}
=\frac{\Lambda}{3}(1-\gamma^{2}\xi^{2}\frac{\Lambda}{3}).
\label{deS}
\end{equation}

In this special case $\alpha=0$ and $n=0$ we can integrate eq.(1) for
$\Lambda>0$ and choosing the integration constant equal to zero
\begin{equation}
p(t) = \text{sgn}(p) \frac{\Lambda_{0,\mu_0}}{\Lambda} \frac{1}{ \cosh^{2}{\left(\sqrt{\frac{\Lambda}{3}}t \right)} },
\end{equation}
which gives the maximal value o parameter $p$  
\begin{equation}
|p_{\text{max}}| = \frac{\Lambda_0}{\Lambda}. 
\end{equation}
In the same case with $\Lambda<0$ and choosing integration constant equal to zero we have 
\begin{equation}
p(t) = \text{sgn}(p) \frac{\Lambda_{0,\mu_0}}{\Lambda} \frac{1}{ \sinh^{2}{\left(\sqrt{\frac{\Lambda}{3}}t \right)} }.
\end{equation}

For $\alpha=0$ and $\Lambda<\Lambda_{0,\bar{\mu}}$ in the case $n=-1/2$ we
obtain the de Sitter solution
\begin{equation}
p(t) = \text{sgn}(p) \ \text{const} \cdot \exp{\left\{ 2\sqrt{1-\frac{\Lambda}{\Lambda_{0,\bar{\mu}}} } \sqrt{\frac{\Lambda}{3}}t \right\} }. 
\end{equation}

These solutions represent the lines on the boundaries of the physically admissible domains.

\section{Summary} \label{sec:summary}

We have studied dynamics and analytical solutions of the flat Friedmann-Robertson-Walker
cosmological model with a free scalar
field and the cosmological constant, modified by the holonomy corrections of
Loop Quantum Gravity. 

We performed calculations in two setups called
$\bar{\mu}-$scheme and  $\mu_0-$scheme, explained in the appendix A. 
We have explored  whole $\Lambda \in \mathbb{R}$ range and  whole allowed
$p^2_{\phi} \in \mathbb{R}_+ \cup \{ 0\}$ range.  In the case of $\bar{\mu}-$scheme
resulting solutions are of oscillating type for $\Lambda \in \mathbb{R}_-$ and 
bouncing type for $\Lambda \in (0,\Lambda_0)$.  For  $\Lambda \in [\Lambda_0,\infty)$
there are no physical solutions. In the case of $\mu_0-$scheme
for $\Lambda \in \mathbb{R}_-$ bouncing solutions occur. 
When both $\Lambda \in \mathbb{R}_+$ and relation  (\ref{rel}) are fulfilled,
oscillatory behaviour occurs. Otherwise, bouncing solutions appear. 
In all considered cases with $p^2_{\phi} \in \mathbb{R}_+ $ the initial singularity is avoided. 

We have investigated the evolutional paths of the model, from the point of view of
qualitative methods of dynamical systems of differential equations.
We found that in the special case of $n=-1/2$ the boundary trajectory $(\alpha=0)$
approaches the de Sitter state; and demonstrate that in the case of 
positive cosmological constant there are two types of dynamical 
behaviours in the finite domain. For the case of $n=0$ there appear oscillating 
solutions without the initial and final singularities,
and that they change into bouncing for $0> n\geq -1/2$.     

The results of this paper can give helpful background dynamics to study variety 
of physical phenomenas during the bounce epoch in Loop Quantum Cosmology. 
For example, the interesting question of
the fluctuations like gravitational waves\cite{Bojowald:2007cd,Mielczarek:2007zy,Mielczarek:2007wc}
or scalar perturbations \cite{Bojowald:2006tm} during this period. 
We tried to show that numerical calculations can be ``shifted'' one step
further, since the basic model is explicitly solvable, and can be treated as
starting ground for more complex problems, like the above, which cannot be
solved analytically.

\begin{acknowledgments}
Authors are grateful to Orest Hrycyna for discussion.
This work was supported in part by the Marie Curie Actions Transfer of
Knowledge project COCOS (contract MTKD-CT-2004-517186).
\end{acknowledgments}

\appendix 
\section{Flat FRW model and holonomy corrections in Loop Quantum Gravity} \label{Appendix1}
In this appendix, we derive the form of the Hamiltonian (\ref{model}) considered in the paper.

The FRW $k=0$ spacetime metric can be written as
\begin{equation}
ds^2=-N^2(x) dt^2 + q_{ab}dx^adx^b,
\end{equation}
where $N(x)$ is the lapse function and the spatial part of the metric is expressed as 
\begin{equation}
q_{ab}= \delta_{ij} {\omega^i_a} {\omega^j_b}= a^2(t) {^oq}_{ab} = a^2(t)  \delta_{ij}  {^o\omega^i_a}{^o\omega^j_b}.
\end{equation}
In this expression ${^oq}_{ab}$ is fiducial metric and ${^o\omega^i_a}$ are co-triads dual to the triads 
${^oe^a_i}$,  ${^o\omega^i}({^oe_j})=\delta^i_j$   where $^o\omega^i={^o\omega^i_a}dx^a$ and $^oe_i={^oe_i^a}\partial_a$.
From these triads we construct the Ashtekar variables 
\begin{eqnarray}
A^i_a &\equiv& \Gamma^i_a+\gamma K_a^i = \tilde{c} \ {^o\omega^i_a}  , \label{A} \\
E^a_i &\equiv& \sqrt{|\det q|} e^{a}_i = \tilde{p} \sqrt{^oq} \ {^oe^a_i}
\label{E},
\end{eqnarray}  
where 
\begin{eqnarray}
|\tilde{p}| &=& a^2, \\
\tilde{c} &=& \gamma \dot{a}.
\end{eqnarray}
Note that the Gaussian constraint implies that $\tilde{p} \leftrightarrow  -\tilde{p}$
leads to the same physical results.
The factor $\gamma$ is called Barbero-Immirzi parameter.
In the definition (\ref{A}) the spin connection is defined as
\begin{equation}
\Gamma^i_a = -\epsilon^{ijk}e^b_j(\partial_{[a}e^k_{b]}+\frac{1}{2}e^c_k e^l_a
\partial_{[c}e^l_{b]} ),
\end{equation}
and the extrinsic curvature is defined as 
\begin{equation}
K_{ab}=\frac{1}{2N}\left[ \dot{q}_{ab} -2 D_{(a}N_{b)} \right],
\end{equation}
which corresponds to $K^i_a := K_{ab} e^b_i$.

The scalar constraint, in Ashtekar variables, has the form 
\begin{eqnarray}
H_{\rm G} &=&  \frac{1}{16 \pi G} \int_{\Sigma} d^3 x N(x) \frac{E^a_i
E^b_j}{\sqrt{|\mathrm{det} E|}}  \left[  {\varepsilon^{ij}}_k F_{ab}^k  \right.  \nonumber  \\
      &-& \left.   2(1+\gamma^2)  K^i_{[a}  K^j_{b]} \right],
\label{scalar}
\end{eqnarray}
where field strength is expressed as
\begin{equation}
F^k_{ab}=\partial_aA^k_b-\partial_bA^k_a+\epsilon^k_{ij}A^i_aA^j_b.
\label{str}
\end{equation}
With use of  (\ref{A}),(\ref{E}) and (\ref{str}) the Hamiltonian
(\ref{scalar}) assumes the form
\begin{equation}
H_{\text{G}} =  - \frac{3 V_0}{8 \pi G \gamma^2} \sqrt{|\tilde{p}|}
{\tilde{c}}^2,
\label{hami}
\end{equation}
where we have assumed a gauge of $N(x)=1$.
The constant $V_0$ is the volume of the fiducial cell. 
This volume can be chosen arbitrarily.
It is convenient to absorb
the factor $V_0$ by redefinition
\begin{equation}
p=\tilde{p} V_0^{2/3} , \ c=\tilde{c} V_0^{1/3}.
\end{equation}
The holonomy along a curve $\alpha \in \Sigma$ is defined as follows
\begin{equation}
h_{\alpha} = \mathcal{P} \exp \int_{\alpha} \tau_i A^i_a dx^a,
\end{equation}
where $2 i \tau_i= \sigma_i$ and $\sigma_i$ are the Pauli matrices.
From this definition we can calculate holonomy in the direction $^oe^a_i\partial_a$ and the length $\mu V_0^{1/3}$ 
\begin{eqnarray}
h_{i}^{(\mu)} &=& \exp \int_0^{\mu V_0^{1/3} } \tau_i c V_0^{-1/3}  {^o\omega^i_a} dx^a = \exp{\tau_i \mu c }  \nonumber \\
              &=& \mathbb{I}\cos \left( \frac{\mu c}{2}\right)+2\tau_i\sin
              \left( \frac{\mu c}{2}\right),
\label{hol2}
\end{eqnarray}
where we used definition of the Ashtekar variable $A$ (\ref{A}). From such a particular holonomies we can 
construct holonomy along the closed curve $\alpha=\Box_{ij}$. This curve is schematically presented on the 
diagram below.

\begin{center}
\setlength{\unitlength}{0.7cm}
\begin{picture}(10,10)
\thicklines
\put(3,3){\vector(1,0){2}}
\put(5,3){\line(1,0){2}}
\put(7,3){\vector(0,1){2}}
\put(7,5){\line(0,1){2}}
\put(7,7){\vector(-1,0){2}}
\put(5,7){\line(-1,0){2}}
 \put(3,7){\vector(0,-1){2}}
\put(3,5){\line(0,-1){2}}
\put(4.5,2){\vector(1,0){1}}
\put(5.5,8){\vector(-1,0){1}}
\put(8,4.5){\vector(0,1){1}}
\put(2,5.5){\vector(0,-1){1}}
\put(4.5,1.2){$^oe^a_i\partial_a$}
\put(8.3,5){$^oe^a_j\partial_a$}
\put(4.5,8.5){$-{^oe^a_i\partial_a}$}
\put(0.2,5){$-{^oe^a_j\partial_a}$}
\put(4.5,3.4){$h_{i}^{(\mu)}$}
\put(5.8,4.6){$h_{j}^{(\mu)}$}
\put(4.5,6){$h_{i}^{(\mu)-1}$}
\put(3.3,4.6){$h_{j}^{(\mu)-1}$}
\end{picture}
\end{center}

This holonomy can be written as 
\begin{eqnarray}
h_{\Box_{ij}}^{(\mu)} &=& h_{i}^{(\mu)} h_{j}^{(\mu)} h_{i}^{(\mu)-1} h_{j}^{(\mu)-1} \nonumber \\
                      &=& e^{\mu  V_0^{1/3} A_a {^oe^a_i}}e^{\mu V_0^{1/3}  A_a {^oe^a_j}}e^{-\mu V_0^{1/3} 
 A_a {^oe^a_i}}e^{-\mu  V_0^{1/3} A_a {^oe^a_j}} \nonumber \\
                      &=& \exp\left[ \mu^2 V_0^{2/3}  A^l_{a} A^m_{b}{^oe^a_i} {^oe^b_j} [\tau_l,\tau_m] +
 \mathcal{O}(\mu^3) \right] 
  \nonumber   \\
                      &=& \mathbb{I} + \mu^2  V_0^{2/3} F^k_{ab} \tau_k
                      {^oe^a_i} {^oe^b_j} +   \mathcal{O}(\mu^3),
\label{deriv}
\end{eqnarray}
where we have used Baker-Campbell-Hausdorff formula and the fact 
that for flat FRW, field strength (\ref{str}) simplifies to the form
$F^k_{ab}=\epsilon^k_{ij}A^i_aA^j_b$. Now, equation (\ref{deriv}) can be simply rewritten to the form
\begin{equation}
F^k_{ab} = - 2 \lim_{\mu \rightarrow 0}
 \frac{\text{tr}\left[\tau_k \left( h^{(\mu)}_{\Box_{ij}}-\mathbb{I} \right)  \right]}{\mu^2 V_0^{2/3} } 
{^o\omega^i_a}{^o\omega^j_b}.
\label{lim}
\end{equation}
The trace in this equation can be calculated with use of definition (\ref{hol2}) 
\begin{equation}
\text{tr}\left[\tau_k \left( h^{(\mu)}_{\Box_{ij}}-\mathbb{I} \right)  \right] =
 - \frac{\epsilon_{kij}}{2} \sin^2\left(\mu c \right).
\label{tr}
\end{equation}
In Loop Quantum Gravity the limit $\mu \rightarrow 0$ in the formula (\ref{lim}) does not exist because of
existence of the area gap. The area gap corresponds to the minimal quantum of
area $\Delta=2\sqrt{3}\pi\gamma l_{\text{Pl}}^2$, which arises as the first
non-zero eigenvalue of the are operator \cite{Ashtekar:1996eg}.
So instead of the limit in equation (\ref{lim}) we should stop shrinking the
loop at the appropriate 
minimal area $\Delta$. This area corresponds to the area intersected by the loop.
For the holonomy in the direction $^oe^a_i\partial_a$  the area is $\boxdot_i$, as 
explained in the diagram below (where $i=k \wedge j$). 

\begin{center}
\setlength{\unitlength}{0.3cm}
\begin{picture}(12,10)
\thicklines
\put(3,0){\line(4,3){4}}
\put(3,7){\line(4,3){4}}
\put(3,0){\line(0,1){7}}
\put(7,3){\line(0,1){7}}
\put(0,5){\line(1,0){3}}
\put(5,5){\vector(1,0){2}}
\put(7,5){\line(1,0){3}}
\put(11,4.7){$= \ \boxdot_i$}
\put(8,4){$i$}
\put(5,0.5){$j$}
\put(2,3){$k$}
\end{picture}
\end{center}

The limit $\boxdot_i \rightarrow \Delta$ corresponds to
$\mu \rightarrow \bar{\mu}$.
Now, we must connect the area $\boxdot_i$ with the length $\mu$. 
We can choose that area $\boxdot_i$ to correspond to physical 
area $a^2 \mu^2$ or to the area $\mu^2$. 
So in the case $\boxdot_i=V_0^{2/3} a^2 \mu^2 = |p|\mu^2$ we have the limit  
\begin{equation}
\mu \rightarrow \bar{\mu}  = \sqrt{ \frac{\Delta}{|p|}}.
\end{equation}
This approach we call $\bar{\mu}-$scheme. In the case $\boxdot_i =|p_0| \mu^2$
where $p_0$ correspond to the eigenvalue
\begin{equation}
\hat{p} | \mu_0 \rangle =  \mu_0 \frac{8\pi \gamma l^2_{\text{Pl}}}{6} | \mu_0
\rangle,
\end{equation}
and taking the limit
\begin{equation}
\mu \rightarrow \bar{\mu} = \mu_0 = \frac{3\sqrt{3}}{2},
\end{equation}
which we call $\mu_0-$scheme.

In the quantum version, we can combine equations  (\ref{lim}) and  (\ref{tr}), and write 
\begin{equation}
F^k_{ab} = \frac{  \sin^2\left(\bar{\mu} c \right) }{ \bar{\mu}^2 V_0^{2/3} } \epsilon_{kij}  {^o\omega^i_a}{^o\omega^j_b},
\end{equation}
whereas in the classical case we have
\begin{equation}
F^k_{ab} = \frac{c^2}{V_0^{2/3} } \epsilon_{kij} {^o\omega^i_a}{^o\omega^j_b}.
\end{equation}
So, from these two equations we see that quantum effects can be introduced by a replacement
\begin{equation}
c \rightarrow  \frac{ \sin \left( \bar{\mu} c  \right) }{\bar{\mu}} 
\end{equation}
in the classical expressions. The Hamiltonian (\ref{hami}) with holonomy
correction takes the form
\begin{equation}
H_{\text{eff}} =  - \frac{3}{8 \pi G \gamma^2} \sqrt{|p|} \left[ \frac{ \sin \left( \bar{\mu} c
\right) }{\bar{\mu}}   \right]^2 .
\end{equation}


\begin{thebibliography}{99}

%\cite{Singh:2006im}
\bibitem{Singh:2006im}
  P.~Singh, K.~Vandersloot and G.~V.~Vereshchagin,
  %``Non-singular bouncing universes in loop quantum cosmology,''
  Phys.\ Rev.\  D {\bf 74} (2006) 043510
  [arXiv:gr-qc/0606032].
  %%CITATION = PHRVA,D74,043510;%%

%\cite{Ashtekar:2006uz}
\bibitem{Ashtekar:2006uz}
  A.~Ashtekar, T.~Pawlowski and P.~Singh,
  %``Quantum nature of the big bang: An analytical and numerical  investigation.
  %I,''
  Phys.\ Rev.\  D {\bf 73} (2006) 124038
  [arXiv:gr-qc/0604013].
  %%CITATION = PHRVA,D73,124038;%%

%\cite{Stachowiak:2006uh}
\bibitem{Stachowiak:2006uh}
  T.~Stachowiak and M.~Szydlowski,
  %``Exact solutions in bouncing cosmology,''
  Phys.\ Lett.\  B {\bf 646} (2007) 209
  [arXiv:gr-qc/0610121].
  %%CITATION = PHLTA,B646,209;%%

%\cite{Ashtekar:2006rx}
\bibitem{Ashtekar:2006rx}
  A.~Ashtekar, T.~Pawlowski and P.~Singh,
  %``Quantum nature of the big bang,''
  Phys.\ Rev.\ Lett.\  {\bf 96} (2006) 141301
  [arXiv:gr-qc/0602086].
  %%CITATION = PRLTA,96,141301;%%

%\cite{Bojowald:2002gz}
\bibitem{Bojowald:2002gz}
  M.~Bojowald,
  %``Isotropic loop quantum cosmology,''
  Class.\ Quant.\ Grav.\  {\bf 19} (2002) 2717
  [arXiv:gr-qc/0202077].
  %%CITATION = CQGRD,19,2717;%%

%\cite{Bojowald:2003md}
\bibitem{Bojowald:2003md}
  M.~Bojowald,
  %``Homogeneous loop quantum cosmology,''
  Class.\ Quant.\ Grav.\  {\bf 20} (2003) 2595
  [arXiv:gr-qc/0303073].
  %%CITATION = CQGRD,20,2595;%%

%\cite{Ashtekar:2006wn}
\bibitem{Ashtekar:2006wn}
  A.~Ashtekar, T.~Pawlowski and P.~Singh,
  %``Quantum nature of the big bang: Improved dynamics,''
  Phys.\ Rev.\  D {\bf 74} (2006) 084003
  [arXiv:gr-qc/0607039].
  %%CITATION = PHRVA,D74,084003;%%

%\cite{Ashtekar:2006es}
\bibitem{Ashtekar:2006es}
  A.~Ashtekar, T.~Pawlowski, P.~Singh and K.~Vandersloot,
  %``Loop quantum cosmology of k = 1 FRW models,''
  Phys.\ Rev.\  D {\bf 75} (2007) 024035
  [arXiv:gr-qc/0612104].
  %%CITATION = PHRVA,D75,024035;%%

%\cite{Bojowald:2006gr}
\bibitem{Bojowald:2006gr}
  M.~Bojowald,
  %``Large scale effective theory for cosmological bounces,''
  Phys.\ Rev.\  D {\bf 74} (2007) 081301
  [arXiv:gr-qc/0608100].
  %%CITATION = PHRVA,D74,081301;%%

%\cite{Bojowald:2008pu}
\bibitem{Bojowald:2008pu}
  M.~Bojowald,
  %``Quantum nature of cosmological bounces,''
  arXiv:0801.4001 [gr-qc].
  %%CITATION = ARXIV:0801.4001;%%

%\cite{Vandersloot:2006ws}
\bibitem{Vandersloot:2006ws}
  K.~Vandersloot,
  %``Loop quantum cosmology and the k = -1 RW model,''
  Phys.\ Rev.\  D {\bf 75} (2007) 023523
  [arXiv:gr-qc/0612070].
  %%CITATION = PHRVA,D75,023523;%%

%\cite{Szulc:2006ep}
\bibitem{Szulc:2006ep}
  L.~Szulc, W.~Kaminski and J.~Lewandowski,
  %``Closed FRW model in loop quantum cosmology,''
  Class.\ Quant.\ Grav.\  {\bf 24} (2007) 2621
  [arXiv:gr-qc/0612101].
  %%CITATION = CQGRD,24,2621;%%

%\cite{Szulc:2007uk}
\bibitem{Szulc:2007uk}
  L.~Szulc,
  %``Open FRW model in Loop Quantum Cosmology,''
  arXiv:0707.1816 [gr-qc].
  %%CITATION = ARXIV:0707.1816;%%

%\cite{Chiou:2007mg}
\bibitem{Chiou:2007mg}
  D.~W.~Chiou,
  %``Effective Dynamics, Big Bounces and Scaling Symmetry in Bianchi Type I Loop
  %Quantum Cosmology,''
  arXiv:0710.0416 [gr-qc].
  %%CITATION = ARXIV:0710.0416;%%

%\cite{Chiou:2007dn}
\bibitem{Chiou:2007dn}
  D.~W.~Chiou,
  %``Effective dynamics for the cosmological bounces in Bianchi type I loop
  %quantum cosmology,''
  arXiv:gr-qc/0703010.
  %%CITATION = GR-QC/0703010;%%

%\cite{Bojowald:2007yy}
\bibitem{Bojowald:2007yy}
  M.~Bojowald,
  %``The Dark Side of a Patchwork Universe,''
  Gen.\ Rel.\ Grav.\  {\bf 40}, 639 (2008)
  [arXiv:0705.4398 [gr-qc]].
  %%CITATION = GRGVA,40,639;%%

%\cite{Dabrowski:2004hx}
\bibitem{Dabrowski:2004hx}
  M.~P.~Dabrowski and T.~Stachowiak,
  %``Phantom Friedmann cosmologies and higher-order characteristics of
  %expansion,''
  Annals Phys.\  {\bf 321} (2006) 771
  [arXiv:hep-th/0411199].
  %%CITATION = APNYA,321,771;%%

%\cite{Ashtekar:1997yu}
\bibitem{Ashtekar:1997yu}
  A.~Ashtekar, J.~Baez, A.~Corichi and K.~Krasnov,
  %``Quantum geometry and black hole entropy,''
  Phys.\ Rev.\ Lett.\  {\bf 80} (1998) 904
  [arXiv:gr-qc/9710007].
  %%CITATION = PRLTA,80,904;%%

%\cite{Domagala:2004jt}
\bibitem{Domagala:2004jt}
  M.~Domagala and J.~Lewandowski,
  %``Black hole entropy from quantum geometry,''
  Class.\ Quant.\ Grav.\  {\bf 21} (2004) 5233
  [arXiv:gr-qc/0407051].
  %%CITATION = CQGRD,21,5233;%%

%\cite{Meissner:2004ju}
\bibitem{Meissner:2004ju}
  K.~A.~Meissner,
  %``Black hole entropy in loop quantum gravity,''
  Class.\ Quant.\ Grav.\  {\bf 21} (2004) 5245
  [arXiv:gr-qc/0407052].
  %%CITATION = CQGRD,21,5245;%%

%\cite{Bojowald:2007cd}
\bibitem{Bojowald:2007cd}
  M.~Bojowald and G.~M.~Hossain,
  %``Loop quantum gravity corrections to gravitational wave dispersion,''
  arXiv:0709.2365 [gr-qc].
  %%CITATION = ARXIV:0709.2365;%%

%\cite{Mielczarek:2007zy}
\bibitem{Mielczarek:2007zy}
  J.~Mielczarek and M.~Szydlowski,
  %``Relic gravitons as the observable for Loop Quantum Cosmology,''
  Phys.\ Lett.\  B {\bf 657} (2007) 20
  [arXiv:0705.4449 [gr-qc]].
  %%CITATION = PHLTA,B657,20;%%

%\cite{Mielczarek:2007wc}
\bibitem{Mielczarek:2007wc}
  J.~Mielczarek and M.~Szydlowski,
  %``Relic gravitons from super-inflation,''
  arXiv:0710.2742 [gr-qc].
  %%CITATION = ARXIV:0710.2742;%%

%\cite{Bojowald:2006tm}
\bibitem{Bojowald:2006tm}
  M.~Bojowald, H.~H.~Hernandez, M.~Kagan, P.~Singh and A.~Skirzewski,
  %``Hamiltonian cosmological perturbation theory with loop quantum gravity
  %corrections,''
  Phys.\ Rev.\  D {\bf 74} (2006) 123512
  [arXiv:gr-qc/0609057].
  %%CITATION = PHRVA,D74,123512;%%

%\cite{Bojowald:2006qu}
\bibitem{Bojowald:2006qu}  M.~Bojowald,
  %``Loop quantum cosmology and inhomogeneities,''
  Gen.\ Rel.\ Grav.\  {\bf 38} (2006) 1771
  [arXiv:gr-qc/0609034].
  %%CITATION = GRGVA,38,1771;%%

%\cite{Ashtekar:1996eg}
\bibitem{Ashtekar:1996eg}  A.~Ashtekar and J.~Lewandowski,
  %``Quantum theory of geometry. I: Area operators,''
  Class.\ Quant.\ Grav.\  {\bf 14} (1997) A55
  [arXiv:gr-qc/9602046].
  %%CITATION = CQGRD,14,A55;%%

\end{thebibliography}
\end{document}